\documentclass[12pt,fleqn,oneside]{book}

\usepackage{graphicx}
\usepackage{epsf,epsfig,graphics}
\usepackage{float}
\usepackage{makeidx}
\usepackage{latexsym}
\usepackage{amssymb}
\newcommand{\BE}{\begin{equation}}
\newcommand{\EE}{\end{equation}}
\newcommand{\BA}{\begin{array}}
\newcommand{\EA}{\end{array}}
\newcommand{\BEA}{\begin{eqnarray}}
\newcommand{\EEA}{\end{eqnarray}}

\begin{document}
\frontmatter
{\topskip 120pt%
\begin{center}
{ \textbf{Adhesion-induced Phase Separation of
Biomembranes--Effective Potential and Simulations} }
 \vspace{36pt}

{ Jia-Yuan Wu}

\emph{Department of Physics and Center for Complex Systems, \\
National Central University, \\
Chungli, 32054 \\
Taiwan}

\date{\today}

\vspace{32pt}

{September, 2005}
\end{center}}
\thispagestyle{empty}

\chapter*{Abstract}
We present theoretical analyses and numerical simulations for the
adhesion-induced phase separation of multi-component membranes with
two types of ligand-receptor complexes (junctions). We show that
after integrating all possible distributions of the junctions, the
system can be regarded as a membrane under an effective external
potential. Mean field theory and Gaussian approximation are used to
analyze the effective membrane potential and we find (i) The height
difference of the junctions is the main factor that drives phase
separation at sufficiently large junction height difference. (ii) In
the two phase region far from the mean-field critical point, because
of the higher entropy associated with the softer junctions, phase
coexistence occurs when the effective binding energy of the more
rigid junctions is higher. (iii) In the two phase region near the
mean-field critical point, the shape of the effective potential
shows that the phase coexistence occurs when the effective binding
energy of softer junctions is higher. The effect of junction density
on the critical point is studied by Monte Carlo simulations, and the
result shows that phase separation occurs at larger junction height
difference as junction density of the system decreases.
\thispagestyle{empty}

\begin{center}
\tableofcontents%


\end{center}

\newpage
{\topskip 84pt%
{\huge \bf Notation} \\

\begin{table}[h]
\begin{center}
\begin{tabular}{lll}
{\large \emph{Material parameters}} & & \\ \\
symbol & dimensionless form & physical meaning \\
$\kappa$ & & bending rigidity  \\
$\gamma$ & $\frac{\gamma_\alpha a^2}{\kappa}$ & surface tension   \\
$\lambda_\alpha$ & $\frac{\lambda_\alpha a^2}{\kappa}$ & elastic constant of type-$\alpha$ junction \\
$h_\alpha$ & $\ell_\alpha$ &  natural length of type-$\alpha$ junction \\
$E_{B\alpha}$ & $\tilde E_{B\alpha}$ & binding energy of type-$\alpha$ junction \\
$\Phi_\alpha$ & $\phi_\alpha$ & density of type-$\alpha$ junction\\
$\mu_{\alpha}$ & $\tilde \mu_{\alpha}$ & chemical potential of type-$\alpha$ junction \\
$a$ & & lattice constant in the $x-y$ plane\\
$h_0$ & & unit length in the $z$-direction \\
$k_BT$ & & energy unit
\end{tabular}%
\end{center}
\label{notation1}
\end{table}

\begin{table}[h]
\begin{center}
\begin{tabular}{lll}
{\large  \emph{Defined parameters}} \\ \\
$\ell_0$ = $\frac{\ell_1+\ell_2}{2}$ & the average height of the two junctions  \\
$\Delta_h$ = $\frac{\ell_2-\ell_1}{2}$ & length difference of the two junctions \\
$\Lambda_\alpha$ = $\frac{\lambda_\alpha a^2}{\kappa}$ & dimensionless junction elastic constant \\
$\Lambda_+$ = $\Lambda_1+\Lambda_2$ & sum of the junction elastic constant \\
$\Lambda_-$ = $\Lambda_1-\Lambda_2$ & difference of the junction elastic constant \\
$\lambda$ = $\frac{\Lambda_-}{\Lambda_+}$ & junction flexibility difference \\
$E_{eff_\alpha}$ = $\tilde E_{B_\alpha}$+$\tilde \mu_\alpha$ & effective binging energy \\
$E_{eff+}$ = $E_{eff_1}+E_{eff_2}$ & sum of the effective binding energy \\
$E_{eff-}$ = $E_{eff_1}-E_{eff_2}$ & difference of the effective binding energy \\
$g$ = ${\Delta_h}^2\Lambda_+$ & junction height difference \\
\end{tabular}%
\end{center}
\label{notation2}
\end{table}
}

\mainmatter
\chapter{Introduction \label{chapter 1}}
\pagenumbering{arabic}%
\setcounter{page}{1}%
Membrane adhesion has been studied theoretically and experimentally
[1]-[11]. It plays an important role in many biological processes
ranging from embryological development, repair of tissue, immune
response \cite{ref:exp}, to pathology of tumors \cite{ref:alberts_book}. In
general, the adhesion is mediated by some kinds of lock-and-key
molecular complexes (for simplicity, they are called junctions). An
example of membrane adhesion is the formation of immunological
synapse between a T lymphocyte (T cell) and an antigen-present cell
(APC), a key event of the immune response \cite{ref:exp}. In this
example a highly organized pattern in membrane adhesion region where
the TCR/MHC-peptide complexes aggregate in the center with a
peripheral ring composed of the LFA-1/ICAM-1 complexes is observed.

Qi et. al. \cite{ref:Qi_pnas_01} studied the immunological synapse
pattern formation with a theoretical model based on a
Landau-Ginzburg free energy and a set of coarse-grained
reaction-diffusion equations and suggested that the dynamics of
immunological synapse pattern formation is a spontaneous
self-assembly process. Later, Raychaudhuri et. al.
\cite{ref:Raychaudhuri_PRL_03} developed an effective membrane model
which is derived from the reaction-diffusion equations in
\cite{ref:Qi_pnas_01} and the condition of forming immunological
synapse is studied by the mean field, Gaussian, and renormalization
group theories.

In \cite{ref:lipowsky_EPL_02} Lipowsky et. al. studied membrane
adhesion in the presence of one type of junctions and one type of
repellers (repulsive molecules) in their Monte Carlo Simulation. The
results show that the phase separation depends on the height
difference and the concentration of two types of molecules. Later in
\cite{ref:lipowsky_BPJ_04}, they considered membrane adhesion in the
presence of two different types of junctions which resembles T
cell-APC adhesion. Their study indicated that the phase separation
is driven by the height difference between two types of junctions,
but the formation of target-patterned immunological synapse has to
be assisted by the motion of cytoskeleton.

We study the general case of two membranes binding to each other due
to the presence of two types of junctions. Similar system has been
studied in \cite{ref:Chen_PRE_03}, where Chen considered the effect
of junction flexibility difference on phase separation and developed
an equilibrium statistical mechanical analysis which provided a
phase diagram in mean field level. In this thesis, we re-examine the
system studied in \cite{ref:Chen_PRE_03} with an effective membrane
model that is closely related to \cite{ref:Raychaudhuri_PRL_03}. Our
goal is to provide a complete picture for the phase separation
induced by membrane adhesion which includes the effect of junction
height difference, junction flexibility difference, and thermally
activated membrane height fluctuations. This thesis is organized as
follows. Chapter 2 introduces the model and the effective potential
approach, meanwhile mean field theory and Gaussian approximation are
introduced to study the phase diagrams of the system. Chapter 3
discusses the simulation. Chapter 4 summarizes this thesis.

\setcounter{equation}{0}
\chapter{The Model \label{chapter 2}}
In this chapter, we introduce the Hamiltonian of the system under
consideration and derive the effective potential acting on the
membrane. Through the analysis of effective potential in the
mean-field approach and Gaussian approximation, a lot of the physics
of this system are revealed.

\begin{figure}[tbp]
\epsfxsize= 5.5 in \epsfbox{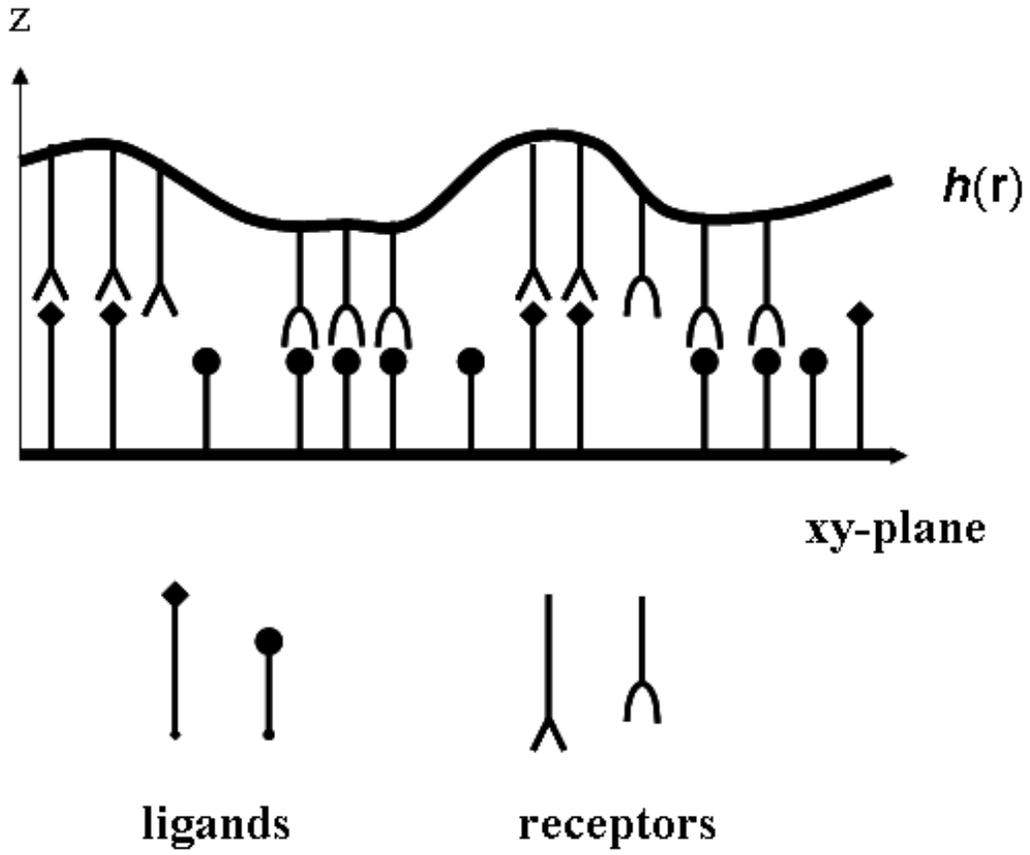} \vskip 24 pt \caption {The
geometry of the system. The system is composed of a
substrate-supported membrane in the bottom and an upper membrane
floating in the solvent. } \vskip 30 pt \label{fig:membrane}
\end{figure}

The main compositions of biomembranes are amphiphilic lipids with
hydrophilic heads and hydrophobic tails. Besides the lipids, there
are many kinds of proteins and carbohydrates. For simplicity, in our
model we consider membranes containing only lipids, ligands, and
receptors. The adhesion between a substrate-supported membrane and a
membrane floating in the solvent is induced by the formation of two
types of specific ligand-receptor complexes (in the rest of this
thesis, we shall refer them as ``junctions''). The geometry of the
system under consideration is shown schematically in Fig.
\ref{fig:membrane}. The height of the upper membrane at $\mathbf{r}$
is denoted as $h(\mathbf{r})$, where $\mathbf{r}=(x,y)$ is a
two-dimensional planar vector. The junctions which mediate the
membrane adhesion are formed by non-covalent adhesion between
type-$\alpha$ ($\alpha$ is 1 or 2) ligands and type-$\alpha$
receptors. The density of type-$\alpha $ junctions at $\mathbf{r} $
is denoted as $\Phi _{\alpha }(\mathbf{r})$, and the binding energy
of a type-$\alpha $ junction is $E_{B\alpha }$. These
ligand-receptor bonds are non-covalent bonds composed of several
hydrogen or van der Waals- like interactions. Typical binding energy
for a ligand-receptor pair is on the order of 10 $k_BT$. Our model
geometry directly applies to the experimental set up in
\cite{ref:exp}. In Appendix A, we show that by suitable
transformation, the system with two membranes in the solvent is
equivalent to to the system described above.
\section{The Hamiltonian}
The energy of this system at any instance has several contributions:
(i)\ the curvature elasticity and surface tension of the membrane,
(ii)\ the binding energy of the ligand-receptor pairs, (iii)\ the
deviation of the ligand-receptor pairs from their natural
conformation. Therefore the simplest coarse-grained Hamiltonian that
includes all the above effects has the following form %
\BEA H &=& \int d^{2}r \left\{ \frac{\kappa}{2} \left[
\nabla^{2}h(\mathbf{r}) \right]^{2} + \frac{\gamma }{2} \left[
\nabla h(\mathbf{r}) \right] ^{2} + \sum_{\alpha=1}^{2}\Phi _{\alpha
}(\mathbf{r})\frac{\lambda _{\alpha }}{2} \left[ h( \mathbf{r}) -
h_{\alpha } \right] ^{2} \right. \nonumber \\ && \left. -
\sum_{\alpha =1}^{2}\Phi_{\alpha}(\mathbf{r})E_{B\alpha }\right\}.
\label{eq:H}
\EEA %
Here $\kappa$ is the bending rigidity of the membrane (typical value
of $\kappa$ is $2 - 70 \times 10^{-20}\ N\cdot m$
\cite{ref:sackmann_BJ_95}). $\gamma$ is the surface tension of the
membrane (typical value of $\gamma$ is $24 \times 10^{-6}\ N/m$
\cite{ref:hochmuth_BPJ_92}. In this simple model it is assumed that
$\kappa$ and $\gamma$ are independent of the composition of the
membranes, and the direct interactions between the junctions, free
ligands, and free receptors are neglected. The nonspecific
interactions between the membranes are also neglected. When a
type-$\alpha$ junction is at $\mathbf{r}$, the interaction energy
between the membranes acquires a minimum at $h=h_{\alpha}$ (the
natural height of a type-$\alpha$ junction, typical value 10 - 30 nm
\cite{ref:bell_BPJ_84}), and the coupling term $\sum _{\alpha =
1}^{2} \Phi_{\alpha}(\mathbf{r}) \frac{\lambda _{\alpha}}{2} \left(
h(\mathbf{r}) - h_{\alpha} \right)^2$ comes from the Taylor
expansion of $h$ around $h _\alpha$, where $\lambda _{\alpha}$ is
the flexibility of a type-$\alpha$ junction (typical value of
$\lambda_\alpha$ is $10^{-5} - 1\ N/m$ \cite{ref:bell_BPJ_84}). The
typical values of the material parameters mentioned above are listed
in Table. \ref{tab:parameters}.
\begin{table}[tbp]
\caption{Typical values of material parameters measured for resting
cells} \vskip 24 pt
\begin{center}
\begin{tabular}{ccc}
\hline \\
parameter  & symbol & typical value  \\ \\
\hline\hline \\
bending rigidity & $\kappa$ & $2 - 70 \times 10^{-20}\ N\cdot m$ \cite{ref:sackmann_BJ_95} \\
surface tension  & $\gamma$ & $24 \times 10^{-6}\ N/m$ \cite{ref:hochmuth_BPJ_92} \\
junction elastic constant & $\lambda_\alpha$ & $10^{-5} - 1\ N/m$ \cite{ref:bell_BPJ_84} \\
junction length  & $h_\alpha$ & $10 - 30$ nm \cite{ref:bell_BPJ_84} \\
junction binding energy   & $E_{B\alpha}$ & $\sim 10\ k_BT$ \\
number of or ligands (receptors)  & & $10^5/\rm{cell}$ \cite{ref:bell_BPJ_84} \\ \\
\hline
\end{tabular}%
\end{center}
\vskip 36 pt \label{tab:parameters}
\end{table}
To proceed our discussion, we put the variables $h(\mathbf{r})$ and
$\Phi _{\alpha }(\mathbf{r})$ on a two-dimensional lattice with
lattice constant $a$. The size of $a$ is chosen to be the smallest
length scale beyond which the continuum elastic description of
membranes breaks down ($a = 6\ \mathrm{nm}$
\cite{ref:lipowsky_PRL_99}). The energy unit is chosen to be
$k_{B}T$, where the unit length  in the $x-y$ plane is $a$ and
$h_{0}= a/\sqrt{\frac{\kappa }{k_{B}T}}$ in the $z$ direction. That
is, $\tilde{E}_{B\alpha }=E_{B\alpha}/k_{B}T$, $\ell
_{i}=h(\mathbf{r})/h_0$ and $\phi_{\alpha }(i) = a^{2}\Phi _{\alpha
}(\mathbf{r})$. So the
non-dimensional form of the Hamiltonian is %
\BEA H_{lat} &=& \sum _{i = 1 }^{N} \left\{ \frac{1}{2}(\Delta
_d \ell _i)^2 + \frac{1}{2} \frac{\gamma
a^2}{\kappa}(\mathbf{\nabla} \ell _i)^2 \nonumber \right. \\ &&
\left. + \sum _{\alpha = 1}^{2} \phi _{\alpha}(i) \left[ \frac{1}{2}
\frac{\lambda _{\alpha} a ^2}{\kappa} \left[ \ell(i) - \ell
_{\alpha} \right]^2 - \tilde E _{B\alpha} \right] \right\}.
\label{eq:non-dim H}
\EEA%
Where the discretized Laplacian $\Delta _{d}$ in two dimension is
given by $\Delta _{d}\ell _{i} = \Delta _{d}\ell _{x,y} = \ell
_{x+1,y}+\ell _{x-1,y}+\ell_{x,y+1}+\ell _{x,y-1}-4\ell _{x,y}$, and
the discretized gradient $\mathbf{\nabla}$ is $\mathbf{\nabla} \ell
_{i} = \mathbf{\nabla }\ell _{x,y} = \frac{1}{2}(\ell _{x+1,y}-\ell
_{x-1,y})\hat{x}+\frac{1}{2}(\ell _{x,y+1}-\ell _{x,y-1})\hat{y}$.
\ Since the calculation in the canonical ensemble (system with fixed
total number of ligands and receptors) is cumbersome. Thus, for
simplicity the analysis is carried out in the grand canonical
ensemble and we introduce the chemical potential of type-$\alpha$
junction, $\mu _\alpha$ ($\mu_\alpha = k_BT \tilde \mu_\alpha$).
Here we apply the hard-core repulsion between the junctions,
consequently each site can be occupied by lipids or one type-1
junction or one type-2 junction. Therefore, the partition function
of the system can be written as \BEA Z &=&\int \mathcal{D}[\ell ]\
e^{-H_{e\ell }[\ell ]} \prod_{i=1}^{N}\left[ 1 + e^{ E_{eff1}
-\frac{1}{2}\frac{\lambda a^2}{\kappa }(\ell_{i}-\ell_1)^2 } + e^{
E_{eff2} - \frac{1}{2}\frac{\lambda a^2}{\kappa }(\ell_{i}-\ell_2)^2
} \right]
\label{eq:partition fun} \\
&=&\int \mathcal{D}[\ell ]\ e^{-(H_{e\ell}+U_{eff})}, \EEA%
where $E_{eff\alpha} \equiv \tilde{E}_{B\alpha } +
\tilde{\mu}_{\alpha }$ is the effective binging energy which
represents the combinations of binding energy of ligand-receptor
pairs and entropy lost of free ligands and free receptors and  \BEA
H_{e\ell } = \sum_{i=1}^{N}\left[\frac{1}{2}(\Delta_{d}\ell_{i})^{2}
+\frac{1}{2}\frac{ \gamma a^{2}}{\kappa
}(\mathbf{\nabla}\ell_{i})^{2}\right] \label{eq:Hel} \EEA %
is the elastic energy of the membrane and \BEA U_{eff} =
-\sum_{i=1}^{N}\ln\left[1+\sum_{\alpha=1}^{2}
e^{E_{eff\alpha}-\frac{1}{2}\frac{\lambda a^2}{\kappa }
(\ell_{i}-\ell_{\alpha })^{2}}\right] \equiv \sum_{i=1}^{N}
V_{eff}(i) \label{eq:effective potential}
\EEA%
is the effective potential acting on the membrane by summing over
all possible distributions of junctions. $V_{eff}(i)$ is the
effective potential at site $i$. The total number of type-$\alpha$
junctions $N_\alpha$ in the system can be expressed as the
derivative of the
Grand potential with respect to the chemical potential.%
\BEA G = -k_BT\ln Z,\ N_\alpha = - \frac{\partial G}{\partial
\mu_\alpha} = \sum_i \langle
\frac{e^{E_{eff\alpha}-\frac{1}{2}\frac{\lambda_\alpha a^2}{\kappa }
(\ell_{i}-\ell_{\alpha })^{2}}}{ 1 + \sum_\beta
e^{E_{eff\beta}-\frac{1}{2}\frac{\lambda_\beta a^2}{\kappa}
(\ell_{i}-\ell_{\beta })^{2}}}\rangle. \label{eq:number of
junctions}
\EEA%
Since typical experiments (either living cells or artificial
membranes) are carried out for systems with fixed total number of
ligands and receptors. At the end of our analysis we connect
$E_{eff\alpha}$ to the corresponding junction binding energies and
densities of ligands and receptors in this system, and compare these
values with typical biological systems .
\section{Phase Diagram: Zero Fluctuation }
\begin{table}[h]
\caption{Dimensionless parameters introduced in Sec. 2.2} \vskip24pt
\begin{center}
\begin{tabular}{ccll}
\\ \hline\hline\\%
$\ell_0$            =  $\frac{\ell_1+\ell_2}{2}$     \\
$\Delta_h$       =  $\frac{\ell_2-\ell_1}{2}$     \\
$\Lambda_\alpha$ =  $\frac{\lambda_\alpha a^2}{\kappa}$  \\
$\Lambda_+$      =  $\Lambda_1+\Lambda_2$, $\Lambda_-$ = $\Lambda_1-\Lambda_2$ \\
$\lambda$        =  $\frac{\Lambda_-}{\Lambda_+}$ \\
$E_{eff+}$       =  $E_{eff_1}+E_{eff_2}$, $E_{eff-}$ =  $E_{eff_1}-E_{eff_2}$ \\
$g$              = ${\Delta_h}^2\Lambda_+$ \\
\\ \hline
\end{tabular}
\end{center}
\vskip 36 pt \label{tab:new parameters}
\end{table}
As has been shown in the previous section, the equilibrium membrane
height conformation of the system is the same as that of a membrane
with effective Hamiltonian \BEA
H_{eff} &=& H_{e\ell} + U_{eff} \label{eq:Heff}\\
&=& \sum_{i=1}^{N}\left[\frac{1}{2}(\Delta_{d}\ell_{i})^{2}
+\frac{1}{2}\frac{ \gamma a^{2}}{\kappa
}(\mathbf{\nabla}\ell_{i})^{2} -\ln\left(1+\sum_{\alpha=1}^{2}
e^{E_{eff\alpha}-\frac{1}{2}\Lambda_\alpha (\ell_{i}-\ell_{\alpha
})^{2}}\right) \right]. \nonumber
\EEA%
For convenience, several dimensionless parameters are defined in
Table. \ref{tab:new parameters}. Substitute these parameters into
Eq. (\ref{eq:effective potential}), $V_{eff}({i})$ can be expressed
as a function of $\frac{\ell_i-\ell_0}{\Delta_h} \equiv z$, \BEA
V_{eff}(i) = -\ln \left[ 1 + e^{ \frac{E_{eff+}-g/2}{2} }A(z)
\right], \label{eq:Veff in different type}\EEA %
where \BEA A(z) = e^{\frac{E_{eff-}-g\lambda/2}{2} - g\frac{1 +
\lambda}{4} \left[ (z + 1)^2 -1 \right]} +
e^{-\frac{E_{eff-}-g\lambda/2}{2}- g\frac{1-\lambda}{4} \left[ (z -
1)^2 -1 \right]}. \label{eq:A of z} \EEA The equilibrium membrane
height is determined by minimizing $V_{eff}(i)$ with respect to $z$,
i.e., \BEA \frac{dV_{eff}(i)}{dz} = -\frac{ e^{
\frac{E_{eff+}-g/2}{2} }}{1 + e^{ \frac{E_{eff+}-g/2}{2} }A(z)}
\frac{d A(z)}{dz} = 0. \label{eq:dV/dz}
\EEA%
From Eq. (\ref{eq:dV/dz}), phase separation occurs when
$\frac{dV_{eff}(i)}{dz}=0$ has three roots (one maximum and two
degenerate minimums). Evidently, the value of $z$ that minimizes
$V_{eff}(i)$ is independent of $E_{eff+}$. Therefore, the phase
boundary is also independent of $E_{eff+}$, and only depends on
$\lambda$, $g$, and $E_{eff-}$. However, it will become clear in the
next section that this only holds in the mean field theory.
\subsection{Symmetric Case} \label{subsec:symmetric Case}
\begin{figure}[tbp]
\epsfxsize= 5.5 in \epsfbox{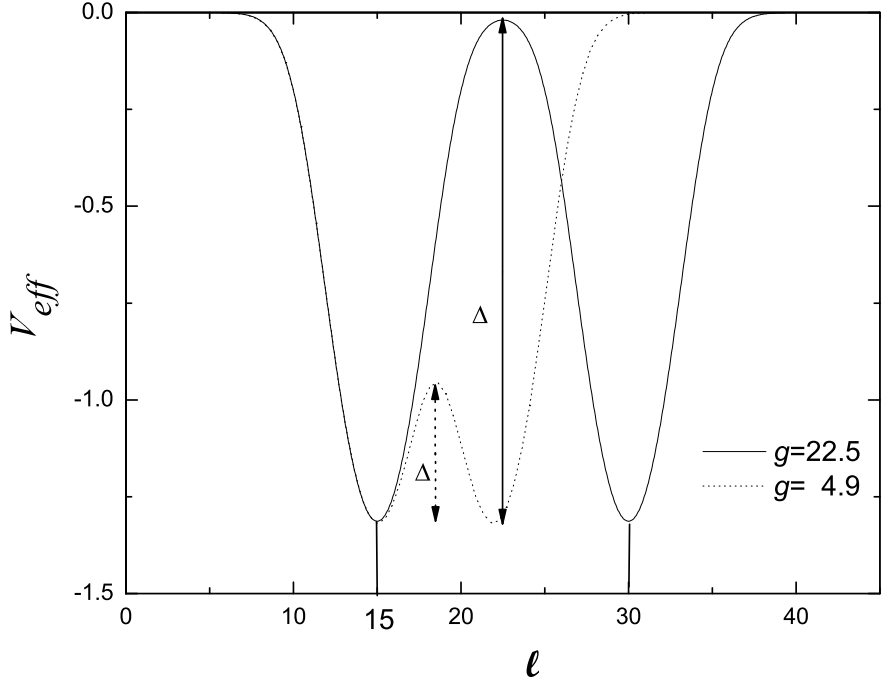} \vskip 24 pt \caption {The
shape of $V_{eff}(i)$ (effective potential) as a function of $\ell$
for two typical cases when phase separation occurs. The parameters
for both lines are $\Lambda_1 =\Lambda_2 = 0.2$, $E_{eff1} =
E_{eff2} = 1$, $\ell_1 = 15$, for solid line $\ell_2 = 30$
($g=22.5$), for dashed line $\ell_2 = 22$ ($g=4.9$). The barrier
height between the two minimums of $V_{eff}(i)$ is denoted as
$\Delta$ which is non-zero when phase separation occurs. The
minimums are near $\ell_1$ and $\ell_2$, the maximum is near $\ell_0
= (\ell_1 + \ell_2)/2$. It also shows that $\Delta$ becomes smaller
when $\ell_2-\ell_1$ decreases. } \vskip 30 pt \label{fig:Veff1}
\end{figure}
\begin{figure}[tbp]
\epsfxsize= 6 in \epsfbox{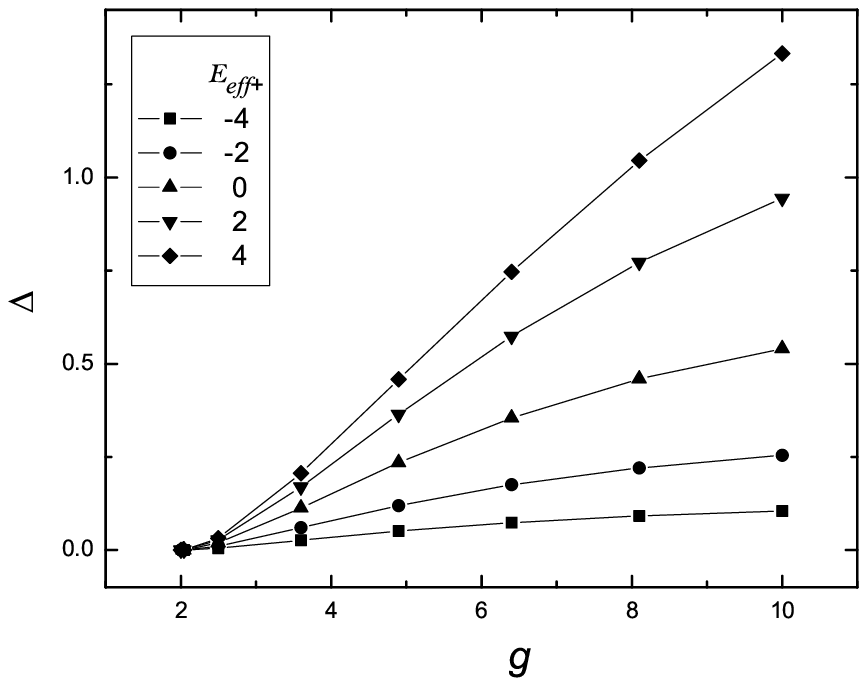} \vskip 24 pt \caption {The
$\Delta - g$ diagram of symmetric case. For different $E_{eff+}$ (by
setting $E_{eff1}=E_{eff2}$), the location of the critical point is
$g=2$ which is independent of $E_{eff+}$. And $\Delta$ decreases as
$E_{eff+}$ decreases.} \vskip
30 pt \label{fig:symmetry for different Eeff}
\end{figure}
In the symmetric case ($\lambda=0$, i.e., both types of junctions
have the same flexibility), the effective potential becomes \BEA
V_{eff}({i}) = -\ln \left[ 1 + e^{ \frac{E_{eff+}-g/2}{2} } ( e^{
\frac{E_{eff-}}{2} - \frac{g}{4} \left[ (z + 1)^2 -1 \right]} + e^{
-\frac{E_{eff-}}{2} - \frac{g}{4} \left[ (z - 1)^2 -1 \right]} )
\right].
\EEA%
In this case, the minimum of $V_{eff}({i})$ depends on $g$ and
$E_{eff-}$. Since when $E_{eff-} = 0$, \BEA A(z) = e^{ - \frac{g}{4}
\left[ (z + 1)^2 -1 \right]} + e^{ - \frac{g}{4} \left[ (z - 1)^2 -1
\right]} \EEA is an even function of $z$. Phase separation occurs at
sufficiently large $g$, and there are three solutions for
$\frac{dA(z)}{dz}=0$. For small $g$, there is only a triple root at
$z = 0$ which corresponds to $\ell_0 = \frac{1}{2}(\ell_1+\ell_2)$.
And we find that the critical point is located at ($E_{eff-}=0$,\ $g
= 2$). The shape of $V_{eff}(i)$ for two typical cases when phase
separation occurs are shown in Fig. \ref{fig:Veff1}. The parameters
for both lines are $\Lambda_1 =\Lambda_2 = 0.2$, $E_{eff1} =
E_{eff2} = 1$, $\ell_1 = 15$, for solid line $\ell_2 = 30$
($g=22.5$), for dashed line $\ell_2 = 22$ ($g=4.9$). The barrier
height between the two minimums of $V_{eff}(i)$ is denoted as
$\Delta$ which is non-zero when phase separation occurs and becomes
smaller when $\ell_2-\ell_1$ decreases. Numerical calculation leads
to the critical point locates at ($E_{eff-}=0$,\ $g = 2$). This
result agrees what we obtain previously. Furthermore, Fig.
\ref{fig:symmetry for different Eeff} shows that for different
$E_{eff+}$ (by setting $E_{eff1}=E_{eff2}$), the critical value of
$g$ is always 2. It confirms that the critical point in the mean
field theory is independent of $E_{eff+}$. And we find that the
values of $\Delta$ decrease as $E_{eff+}$ decreases. That is, the
depth of the potential minimums becomes shallower for smaller
effective binding energy. From the above analysis, we know that the
phase separation is driven by junction height difference, and this
separation only occurs when $g$ is sufficiently large.
\subsection{Asymmetric Case} \label{subsec:Asymmetric Case}
\begin{figure}[tbp]
\epsfxsize= 6 in \epsfbox{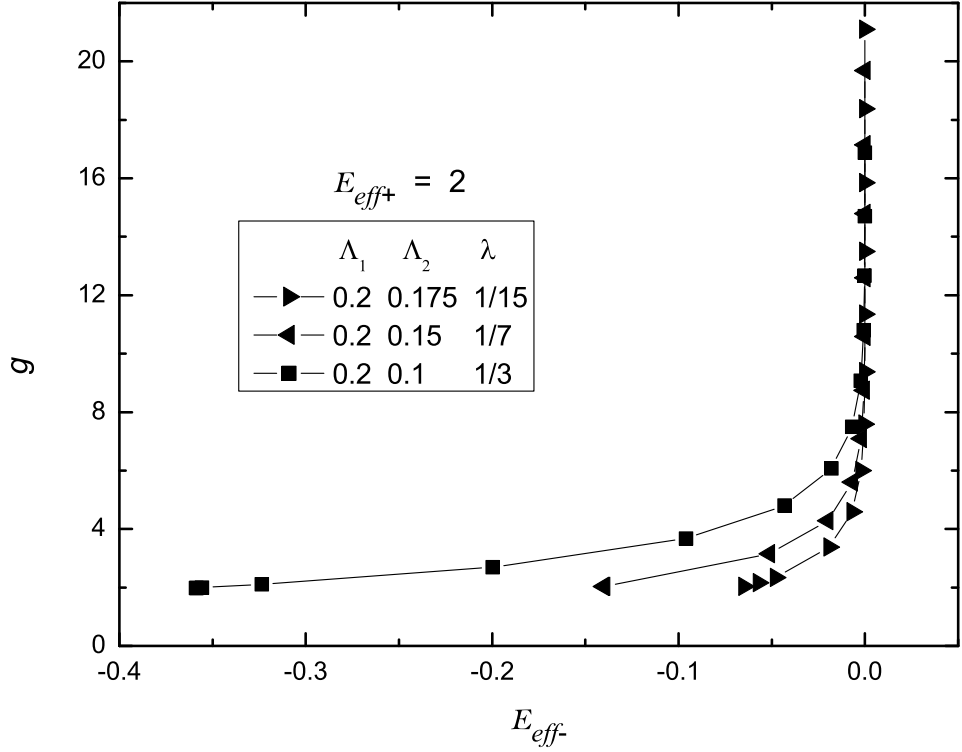} \vskip 24 pt \caption {Mean
field phase boundaries for $E_{eff+}=2$, $\Lambda_1 = 0.2$,
$\Lambda_+ = 0.3$, and $\lambda=1/3$ (square), 1/7
(triangleleft), and 1/15 (triangleright). We find
that near the critical point, phase coexistence curves shift away
from $E_{eff-}=0$ line due to the junction flexibility difference.
Furthermore, at large $g$, the phase boundaries are close to
$E_{eff-}=0$ line.} \vskip 30 pt \label{fig:fix Lambda1}
\end{figure}

\begin{figure}[tbp]
\epsfxsize= 5.5 in \epsfbox{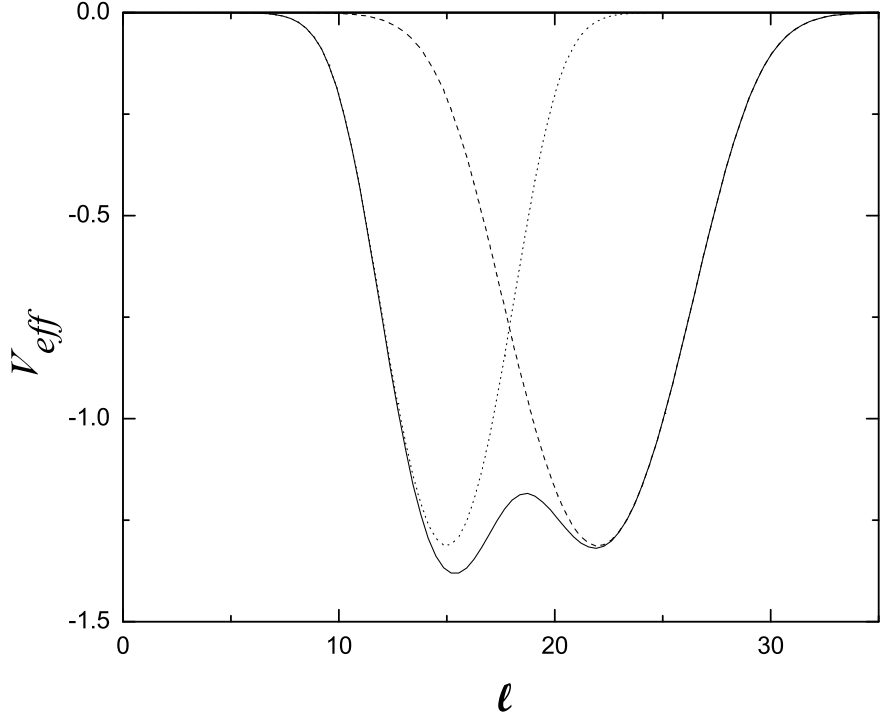} \vskip 24 pt \caption
{Effective potential as a function of $\ell$ in the small $g$ region
with $E_{eff1} = E_{eff2} = 1$, $\Lambda_1=0.2$, $\Lambda_2=0.1$
($\lambda = 1/3$), $\ell_1 = 15$, and $\ell_2 = 22$ ($g=3.657$). The
solid curve is $V_{eff}$, dotted curve for $-\ln \left[ 1 +
e^{E_{eff1} - \frac{\Lambda_1}{2}(\ell - \ell_1)^2} \right]$ which
is the effective membrane potential without the contribution of
type-2 junction, and dashed curve for $-\ln \left[ 1 + e^{E_{eff2} -
\frac{\Lambda_2}{2}(\ell - \ell_2)^2} \right]$ which is the
effective membrane potential without the contribution of type-1
junction. The softer junctions are longer than the more rigid
junctions.  The minimum energy of the more rigid junctions is
``lowered'' by the the softer junctions.} \vskip 30 pt
\label{fig:Veff3}
\end{figure}

Let us consider the more general case $\lambda \neq 0$, i.e.,
different types of junctions have different flexibilities. In this
case, the symmetry in $\lambda=0$ case no longer exists. Therefore
we expect that the phase coexistence curve is no longer located at
$E_{eff-} = 0$ line. An analytical expression of the phase
coexistence curve can be found for $|\lambda| \ll 1$ by expanding
$\frac{d A(z)}{dz}$ around $\lambda = 0$, where $A(z)$ is taken from
Eq. (\ref{eq:A of z}). The critical point for $\lambda \neq 0$
locates at the triple root of $dA(z)/dz = 0$. A straightforward
calculation similar to \cite{ref:Chen_PRE_03} leads to the position
of the critical points for small $\lambda$, \BEA g
&=& 2(1- \frac{\lambda^2}{4}) + \mathcal{O}(\lambda^4), \\
E_{eff-} &=& - \lambda + \mathcal{O}(\lambda^3). \EEA The critical
value of $g$ decreases as the junction flexibility difference
increases, and $E_{eff-}$, $\lambda$ have opposite sign. We present
the phase coexistence curve in two different ways. Fig. \ref{fig:fix
Lambda1+Lambda2} shows phase boundaries for $E_{eff+}=2$, $\Lambda_+
= 0.3$, and $\lambda=1/3$ (square), 4/15 (circle), and 1/6
(triangle). Fig. \ref{fig:fix Lambda1} shows phase
boundaries for $E_{eff+}=2$, $\Lambda_1 = 0.2$, $\Lambda_+ = 0.3$,
and $\lambda=1/3$ (square), 1/7 (triangleleft), and
1/15 (triangleright). Both figures indicate that near the
critical point, phase coexistence curves shift away from
$E_{eff-}=0$ line due to the junction flexibility difference.
Furthermore, at large $g$, the phase boundaries are close to
$E_{eff-}=0$ line. To understand the physics of the phase boundary
in small $g$ region, we plot $V_{eff}(i)$ as a function of $\ell$ in
Fig. \ref{fig:Veff3} with $E_{eff1} = E_{eff2} = 1$,
$\Lambda_1=0.2$, $\Lambda_2=0.1$ ($\lambda = 1/3$), $\ell_1 = 15$,
and $\ell_2 = 22$ ($g=3.657$). The solid curve is $V_{eff}$, dotted
curve for $-\ln \left[ 1 + e^{E_{eff1} - \frac{\Lambda_1}{2}(\ell -
\ell_1)^2} \right]$ which is the effective membrane potential
without the contribution of type-2 junction, and dashed curve for
$-\ln \left[ 1 + e^{E_{eff2} - \frac{\Lambda_2}{2}(\ell - \ell_2)^2}
\right]$ which is the effective membrane potential without the
contribution of type-1 junction. At small $g$, the phase boundary
shifts away from $E_{eff-}=0$ line because the minimum energy of the
more rigid junctions is ``lowered'' by the softer junctions. Thus,
in the region where the height difference is not very large, phase
coexistence occurs when the effective binding energy of the more
rigid junctions is smaller (the effective binding energy of the
softer junctions is larger).
\begin{figure}[tbp]
\epsfxsize= 6 in \epsfbox{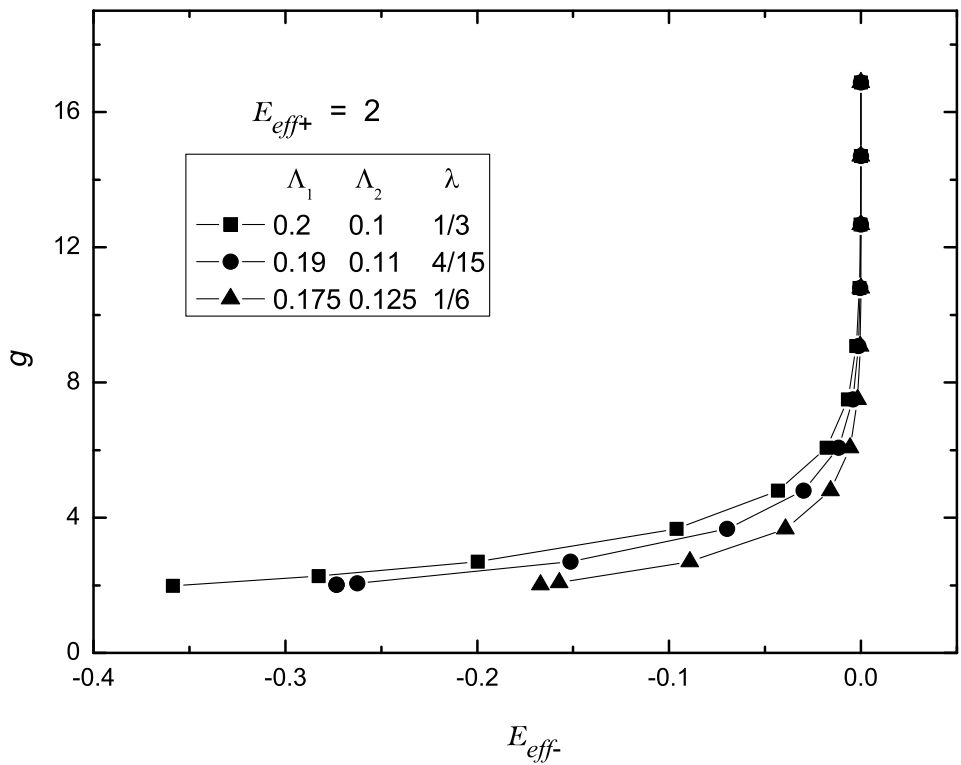} \vskip 24 pt \caption {Mean
field phase boundaries for $E_{eff+}=2$, $\Lambda_+ = 0.3$, and
$\lambda=1/3$ (square), 4/15 (circle), and 1/6
(triangle). We find that near the critical point, phase
coexistence curves shift away from $E_{eff-}=0$ line due to the
junction flexibility difference. Furthermore, at large $g$, the
phase boundaries are close to $E_{eff-}=0$ line.} \vskip 30 pt
\label{fig:fix Lambda1+Lambda2}
\end{figure}
\newpage
\section{Gaussian Approximation}
\begin{figure}[tbp]
\epsfxsize= 5.5 in \epsfbox{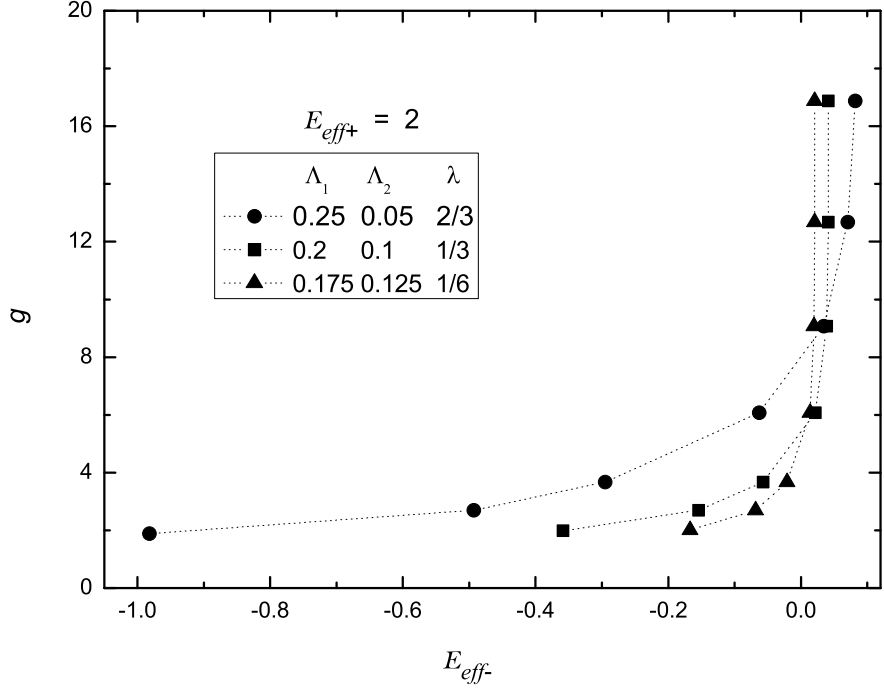} \vskip 24 pt \caption
{Phase boundaries in the Gaussian theory for $E_{eff+}=2$,
$\Lambda_+ = 0.3$, and $\lambda=2/3$ (circle), 1/3
(square), and 1/6 (triangle). Because the softer the
junctions are, the higher the entropy they have, comparing to the
more rigid junctions. The phase boundaries in the large $g$ region
shift away from $E_{eff-}=0$ line (mean field phase boundary), and
the shift is more significant as $|\lambda|$ increases. The critical
points in the figure are taken from the result of the mean field
theory, because the location of critical points are not altered in
Gaussian theory.} \vskip 30 pt \label{fig:GA for different lambda}
\end{figure}

\begin{figure}[tbp]
\epsfxsize= 5.5 in  \epsfbox{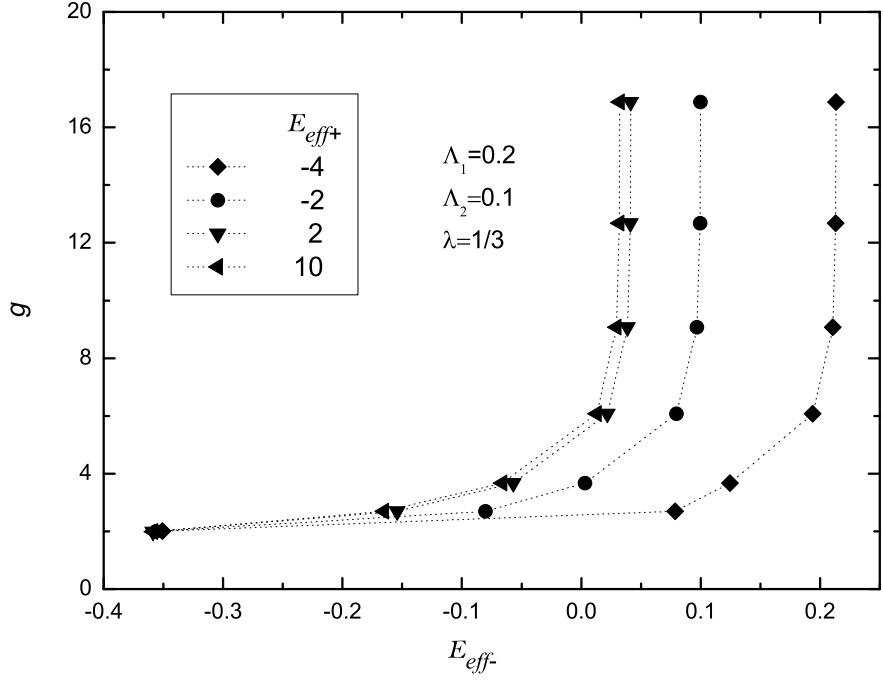} \vskip 24 pt \caption
{Phase boundaries in the Gaussian theory for $\lambda=1/3$,
$E_{eff+}=2$, $\Lambda_+ = 0.3$, and $E_{eff+}=-4$ (lozenge),
-2 (circle), 2 (triangledown), and 10 (triangleleft). The phase boundaries shift away from
$E_{eff-}=0$ line (mean field phase boundary), and the shift is more
significant as $E_{eff+}$ decreases. It is because the entropic
effect of the softer junctions is smaller when the total junction
density is higher (i.e., $E_{eff+}$ is larger). The critical points
in the figure are taken from the result of the mean field theory,
because the location of critical points are not altered in Gaussian
theory.}
 \vskip 30 pt \label{fig:GA for different Eeff}
\end{figure}

In the mean field analysis, we assume that the membrane height is
chosen to minimize $V_{eff}$, phase coexistence occurs when there
are two degenerate minimums, and the critical point is located at
where barrier height between the degenerate minimums vanishes. In this
section, we introduce Gaussian approximation to study the
fluctuation effects on phase boundaries. Since in equilibrium, the
membrane height fluctuates around the potential minimums, the free
energy for a membrane with average height at
$\ell_{\mathrm{min}\alpha}$ (a minimum of $V_{eff}$) in the Gaussian
approximation is calculated by expanding $V_{eff}$ around
$\ell_{\mathrm{min}\alpha}$, i.e., \BEA \frac{F_\alpha}{k_BT} &=& -
\ln \int \mathcal{D}[\ell] e^{ - \left[ H_{el} + \int d^2rV_{eff}
(\ell_{\mathrm{min}\alpha}) + \frac{1}{2} V_{eff}^{\prime\prime}
(\ell_{\mathrm{min}\alpha}) (\ell - \ell_{\mathrm{min}\alpha})^2
\right]}.
\EEA%
Integrating out these Gaussian fluctuations, the free energy of
$\alpha$'th minimum per unit area is \BEA \frac{F_\alpha^*}{k_BT}
&=& V_{eff} (\ell_{\mathrm{min}\alpha}) -
\frac{1}{4\pi}\int_{0}^{(2\pi)^2} dq\ \ln \sqrt{ \frac{2\pi}{q^2 +
\Lambda q + V_{eff}^{\prime\prime} (\ell_{\mathrm{min}\alpha}) } } +
\mathrm{const.}\ \label{eq:free energy}
\EEA%
After comparing the free energy per unit area of the two minimums of
$V_{eff}$, we plot the phase coexistence curves for several choices
of parameters in Fig. \ref{fig:GA for different lambda} and Fig.
\ref{fig:GA for different Eeff}. In large $g$ region, because the
softer junctions can be stretched or compressed easier than the more
rigid junctions and have higher entropy, comparing to mean field
theory predictions, phase coexistence curve moves toward to higher
$E_{eff1}$ ($E_{eff2}$) when $\lambda>0$ ($\lambda<0$). In Fig.
\ref{fig:GA for different lambda}, phase boundaries for several
systems with $E_{eff+}=2$, $\Lambda_+ = 0.3$, and $\lambda=2/3$
(circle), 1/3 (square), and 1/6 (triangle) are
shown. In large $g$ region, because the softer the junctions are,
the higher the entropy they have. Thus, the phase boundary shift is
more significant for systems with lager junction flexibility
difference. Fig. \ref{fig:GA for different Eeff} shows the phase
boundaries for systems with $\lambda=1/3$, $E_{eff+}=2$, $\Lambda_+
= 0.3$, and $E_{eff+}=-4$ (lozenge), -2 (circle), 2
(triangledown), and 10 (triangleleft). In the large
$g$ region, the phase coexistence curves are located farther from
the $E_{eff-}=0$ line (the phase boundary predicted by the mean
field theory) for systems with smaller $E_{eff+}$. It is because the
entropic effect of the softer junctions is smaller when the total
junction density is higher (i.e., $E_{eff+}$ is larger). Fig.
\ref{fig:GA for different lambda} and \ref{fig:GA for different
Eeff} also indicate that in small $g$ region, because the location
of critical points are not altered in the Gaussian theory, therefore
the phase boundary goes to the mean field critical point as $g$
decreases.
\newpage
\section{Summary}
In the mean field analysis, the critical point and phase coexistence
curve is independent of the value of $E_{eff+}$. For symmetric case
 $(\lambda =0)$, the critical point is located at ($E_{eff-}=0,\
g=2)$. The phase coexistence curve lies on the $E_{eff-}=0$ line
starting from $g=2$. In this case, the phase separation is driven by
the height difference of the junctions. For asymmetric case$(\lambda
\neq 0)$, the critical value of $g$ decreases as the junction
flexibility difference increases, and the phase boundaries shift
away from the $E_{eff-}=0$ line in smaller junction height
difference region due to the minimum of the more rigid junction is
``lowered" by the softer junction. In large junction height
difference region, the phase boundaries are close to the
$E_{eff-}=0$ line.

In the Gaussian theory, we find that because the state associated
with membrane height closer to the natural length of the softer
junctions has higher entropy, as a result, comparing to the mean
field predictions, in large $g$ region the phase coexistence curves
in the Gaussian theory move toward higher $E_{eff1}$
($E_{eff2}$)when $\lambda>0$ ($\lambda<0$). The phase boundaries
shift is more significant when the junction flexibility difference
($|\lambda|$) increases or the total junction density ($E_{eff+}$)
decreases. In small $g$ region, because the location of critical
points are not altered in Gaussian theory, thus the phase boundary
goes to the mean-field critical point as $g$ decreases.

\setcounter{equation}{0}
\chapter{Numerical Simulation \label{chapter 3}}
\section{Monte Carlo Simulation}
Monte Carlo simulation is a general name for simulations which use
random sequences. It is a way to perform statistical sampling
experiments on a computer. It is named after the casino city ``Monte
Carlo'' in the Monaco principality.
\subsection{Metropolis Algorithm}
Metropolis algorithm is applied to our simulations. In the
algorithm, if the system is at state $i$ at some instance, first one
randomly chooses another state $j$ as the possible new state of the
system. Let $P_i$ and $P_j$ be the equilibrium probability for state
$i$ and $j$ to occur at temperature $T$. The energy difference
between state $i$ and state $j$ is $\Delta H \equiv E_j - E_i$. If
$\Delta H < 0$, then the system changes its state to state $j$. If
$\Delta H > 0$, then the computer generates a random number
$\mathcal{R}\in[0, 1]$. If%
\BEA \mathcal{R} <\mathcal{S} \equiv \frac{P_j}{P_i} =
e^{-\Delta H} ,\EEA%
then the system changes its state to state $j$, otherwise the system
keeps at state $i$. It can be proved that in the long time limit the
simulation generates a series of states that obey Boltzmann
distribution \cite{ref:binder_book}.
\subsection{Monte Carlo Steps}
In our simulations, the energy of the system depends on the
conformation of the membrane. In each MC step, we repeat the
following procedure $N=L^2$ times. First, randomly choose a site
($i$=($x$, $y$)). The trial conformation of the membrane is the same
as the current state except $\ell_{x,y} \rightarrow
\ell_{x,y}^\prime \equiv \ell_{x,y} + \Delta \ell $, where $-5 <
\Delta \ell <5$. This choice corresponds to a hight change of
maximum size $\sim$ 5 nm. If $\Delta H < 0$, $\ell_{x,y}$ is
replaced by $\ell_{x,y}^\prime$. If $\Delta H > 0$ , then generate a
random number $\mathcal{R}$ and compare to $e^{-\Delta H}$ to decide
whether the height of membrane should be changed or not. We also
choose periodic boundary condition in all our simulations.
\subsection{Snapshots in MC Simulations}
\begin{table}[h]
\caption{Parameters in simulations} \vskip 24 pt
\begin{center}
\begin{tabular}{ccc}
\hline \\
symbol                    & values    \\  \\ \hline\hline \\
$\Gamma$                  &  $0.0025$ \\
$\Lambda_1$, $\Lambda_2$  &  $0.2$    \\
$\ell_1$                  &  $15$     \\ \\
\hline
\end{tabular}%
\end{center}
\vskip 36 pt \label{tab:constant parameters}
\end{table}
The parameters in the simulations are listed in Table.
\ref{tab:constant parameters}. Two snapshots of the contours of the
membrane height with $E_{eff1} = E_{eff2} = 1$, $\Lambda_1 =
\Lambda_2 = 0.2$, $\ell_1 = 15$, and $\ell_2 = 21$ (i.e., $g = 3.6$)
are shown in Fig. \ref{fig:strongly separate snapshot of type-1
junction} and Fig. \ref{fig:strongly separate snapshot of type-2
junction}. The parameters are chosen such that the system is in two
phase region, far from critical region. In Fig. \ref{fig:strongly
separate snapshot of type-1 junction}, the initial membrane height
is set to be $\ell_1$, the natural length of type-1 junction. The
height of the black patches is smaller than $\ell_1$, the height of
the dark gray patches is in the interval between $\ell_1$ and
$\frac{1}{2}(\ell_1 + \ell_2)$. In Fig. \ref{fig:strongly separate
snapshot of type-2 junction}, the initial membrane height is set to
be $\ell_2$, the natural length of type-2 junction. The height of
the light gray patches is in the interval between
$\frac{1}{2}(\ell_1 + \ell_2)$ and $ \ell_2$, the height of the
white patches is larger than $\ell_2$. Both figures show that in two
phase region, the equilibrium membrane height depends on the initial
condition (i.e., the natural length of type-$\alpha$ junction). It
is because the barrier height between the potential minimums is so
large comparing to the thermal fluctuation. Thus, the distribution
of the equilibrium membrane height is near the initial membrane
height.
\begin{figure}[tbp]
\epsfxsize= 3 in \epsfbox{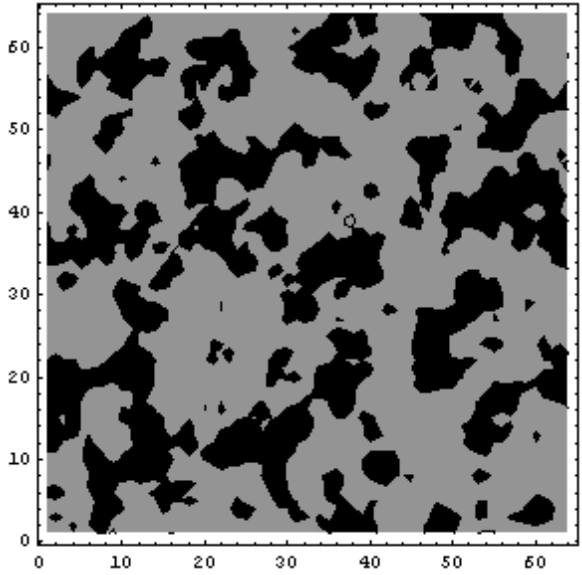} \vskip 24 pt \caption
{Snapshot of the contours of the membrane height with $E_{eff1} =
E_{eff2} = 1$, $\Lambda_1 = \Lambda_2 = 0.2$, $\ell_1 = 15$, and
$\ell_2 = 21$ ($g = 3.6$) on a $64 \times 64$ lattice. The
parameters are chosen such that the system is in two phase region,
far from critical region. The initial membrane height is set to be
$\ell_1$, the natural length of type-1 junction. The height of the
black patches is smaller than $\ell_1$, the height of the dark gray
patches is in the interval between $\ell_1$ and $\frac{1}{2}(\ell_1
+ \ell_2)$. That is, in this region the equilibrium membrane height
is near the natural length of type-1 junction, the initial membrane
height.} \vskip 30 pt \label{fig:strongly separate snapshot of
type-1 junction}
\end{figure}
\begin{figure}[tbp]
\epsfxsize= 3 in \epsfbox{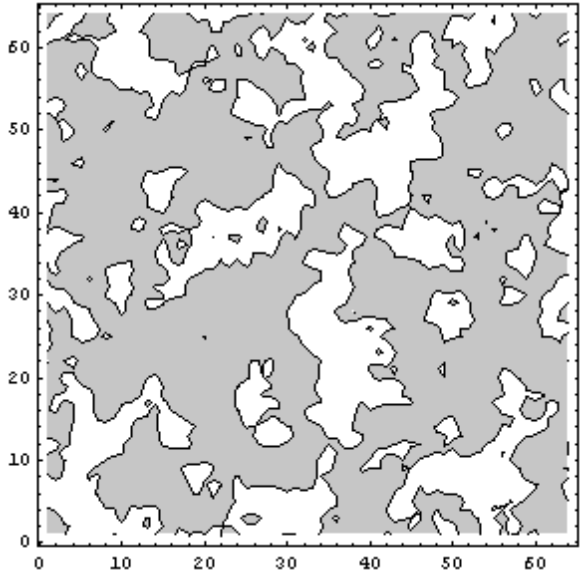} \vskip 24 pt \caption
{Snapshot of the contours of the membrane height with $E_{eff1} =
E_{eff2} = 1$, $\Lambda_1 = \Lambda_2 = 0.2$, $\ell_1 = 15$, and
$\ell_2 = 21$ ($g = 3.6$) on a $64 \times 64$ lattice. The
parameters are chosen such that the system is in two phase region,
far from critical region. The initial membrane height is set to be
$\ell_2$, the natural length of type-2 junction. The height of the
light gray patches is in the interval between $\frac{1}{2}(\ell_1 +
\ell_2)$ and $ \ell_2$, the height of the white patches is larger
than $\ell_2$. That is, in this region the equilibrium membrane
height is near the natural length of type-2 junction, the initial
membrane height.} \vskip 30 pt \label{fig:strongly separate snapshot
of type-2 junction}
\end{figure}
Fig. \ref{fig:weakly separate snapshot of type-1 junction} and Fig.
\ref{fig:weakly separate snapshot of type-2 junction} are snapshots
for $E_{eff1} = E_{eff2} = 1$, $\Lambda_1 = \Lambda_2 = 0.2$,
$\ell_1 = 15$, and $\ell_2 = 20$ ($g = 2.5$) on a $64 \times 64$
lattice. The set of parameters corresponds to a system near critical
point. In Fig. \ref{fig:weakly separate snapshot of type-1
junction}, the initial membrane height is set to be $\ell_1$, the
natural length of type-1 junction. In Fig. \ref{fig:weakly separate
snapshot of type-2 junction}, the initial membrane height is set to
be $\ell_2$, the natural length of type-2 junction. In both figures,
the black, dark gray, light gray, and white color represent the
membrane height $\ell$ correspond to $\ell < \ell_1$, $ \ell_1 <
\ell < \frac{1}{2}(\ell_1 + \ell_2)$, $\frac{1}{2}(\ell_1 + \ell_2)
< \ell < \ell_2$, and $\ell > \ell_2$ respectively. In this region,
the equilibrium membrane height is independent of the initial
condition because the thermal fluctuations make the membrane height
change from shorter (higher) junction phase to higher (shorter)
junction phase.
\begin{figure}[tbp]
\epsfxsize= 3 in \epsfbox{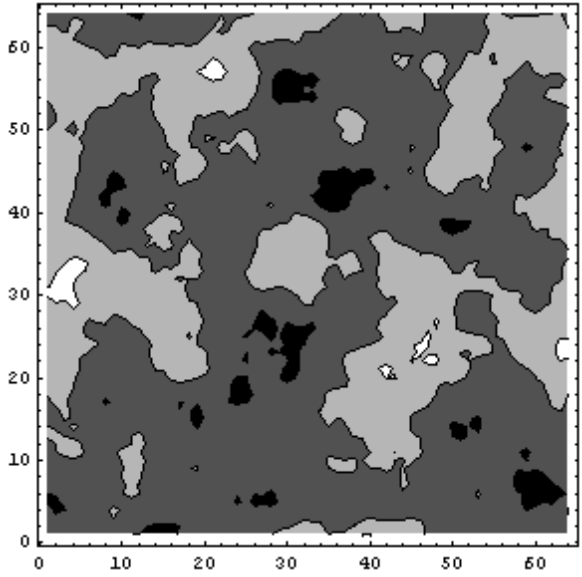} \vskip 24 pt \caption {
Snapshot of the contours of the membrane height with $E_{eff1} =
E_{eff2} = 1$, $\Lambda_1 = \Lambda_2 = 0.2$, $\ell_1 = 15$, and
$\ell_2 = 20$ ($g=2.5$) on a $64 \times 64$ lattice. The parameters
are chosen such that the system is in one phase region, near
critical region. The initial membrane height is set to be $\ell_1$,
the natural length of type-1 junction. The black, dark gray, light
gray, and white color represent the membrane height $\ell$
correspond to $\ell < \ell_1$, $ \ell_1 < \ell < \frac{1}{2}(\ell_1
+ \ell_2)$, $\frac{1}{2}(\ell_1 + \ell_2) < \ell < \ell_2$, and
$\ell > \ell_2$ respectively.} \vskip 30 pt \label{fig:weakly
separate snapshot of type-1 junction}
\end{figure}
\begin{figure}[tbp]
\epsfxsize= 3 in \epsfbox{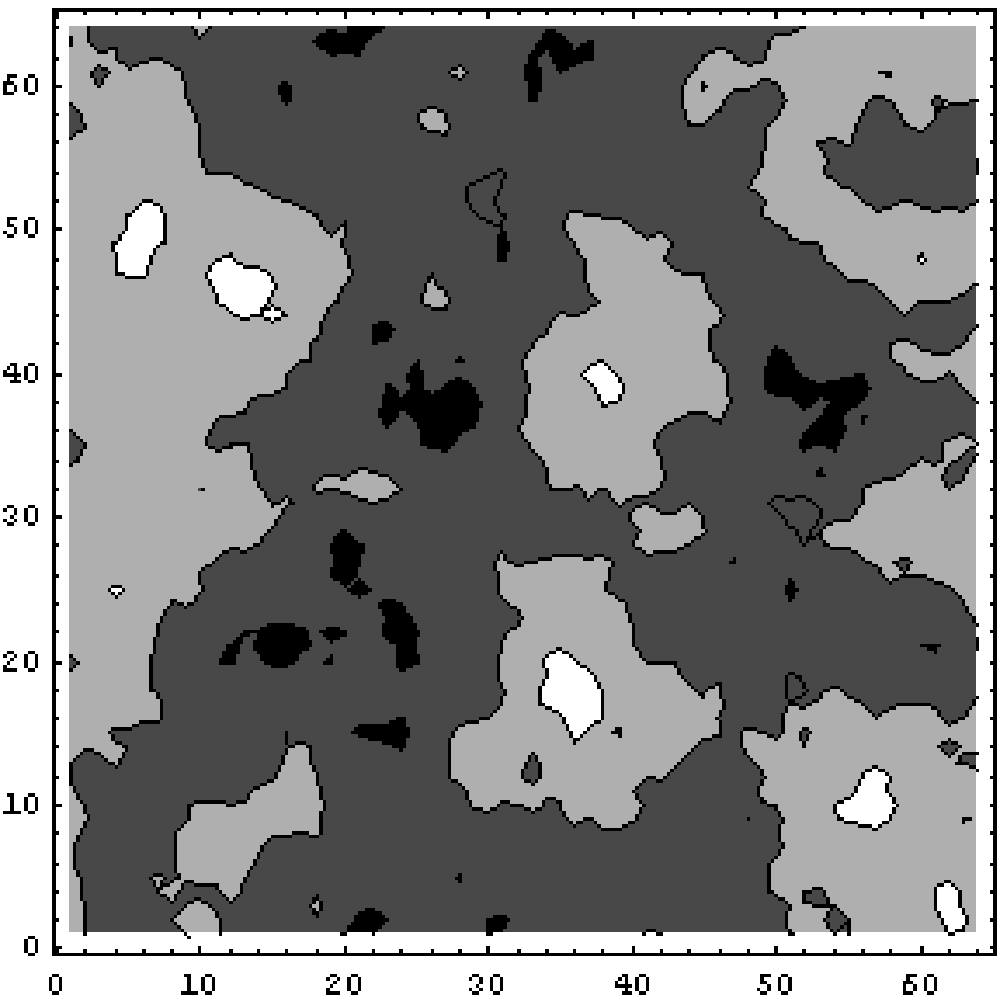} \vskip 24 pt \caption
{Snapshot of the contours of the membrane height with $E_{eff1} =
E_{eff2} = 1$, $\Lambda_1 = \Lambda_2 = 0.2$, $\ell_1 = 15$, $\ell_2
= 20$ ($g=2.5$) on a $64 \times 64$ lattice. The parameters are
chosen such that the system is in one phase region, near critical
region. The initial membrane height is set to be $\ell_2$, the
natural length of type-2 junction. The black, dark gray, light gray,
and white color represent the membrane height $\ell$ corresponds to
$\ell < \ell_1$, $ \ell_1 < \ell < \frac{1}{2}(\ell_1 + \ell_2)$,
$\frac{1}{2}(\ell_1 + \ell_2) < \ell < \ell_2$, and $\ell > \ell_2$
respectively.} \vskip 30 pt \label{fig:weakly separate snapshot of
type-2 junction}
\end{figure}
The above figures show the snapshots of the equilibrium membrane
height for large $g$ (far from the critical region) and small $g$
(near the critical region). However, it is very difficult to locate
the critical point by looking at the snapshots. Thus, we introduce a
systematic method ``Binder cumulant'' to determine the critical
points.

\newpage
\section{The Binder Cumulant and Critical Point}
In last section, we have shown that it is difficult to determine the
critical point quantitatively by looking at the snapshots of the
simulations. Thus, we introduce the Binder cumulants $C_2$ and $C_4$
which are defined as \cite{ref:andelman_EPE_02}:
\BEA%
C_2 = \frac{\langle \bar{z}^2 \rangle}{{\langle \mid \bar{z} \mid
\rangle}^2} \ , \ C_4 = \frac{\langle \bar{z}^4\rangle}{{\langle
\bar{z}^2 \rangle}^2}, \EEA where $\bar{z} = \frac{1}{N}
\sum_{i=1}^{N} z_i$ is the spatial average of the order parameter,
$z_i = \ell_i - \ell_0$, and $\langle \mathcal{O} \rangle$ is the
thermal averages of $\mathcal{O}$. For $g > g_c$ (critical value of
$g$) and $L \gg \xi$ (the correlation length of $z$), the Binder
cumulants reach the values $C_2 =1$ and $C_4 = 1$. For $0 < g < g_c$
and $L \gg \xi$, we have $C_2 = \pi/2 \thickapprox 1.57$ and $C_4 =
3$. For $0 < g < g_c$ and $L \ll \xi$, the moments $C_2$ and $C_4$
vary only weakly with the linear size $L$. Therefore the critical
value of $g$ can be estimated from the common intersection of $C_2$
and $C_4$, respectively, as a function of $g$ for several values of
$L$ \cite{ref:binder_book}, \cite{ref:andelman_EPE_02}.

\section{Results and Discussions}
The number of type-$\alpha$ junctions is determined by
Eq.(\ref{eq:number of junctions}), and we rewrite it as \BEA
N_\alpha = \sum_i \langle \frac{e^{E_{eff\alpha}-\frac{1}{2}
\Lambda_\alpha (\ell_{i}-\ell_{\alpha })^{2}}}{ 1 + \sum_{\beta=1}^2
e^{E_{eff\beta}-\frac{1}{2} \Lambda_\beta (\ell_{i}-\ell_{\beta
})^{2}}}\rangle, \EEA where $\ell_i$ is the membrane height at site
$i$ at some instance. Thus, we can calculate the density of
type-$\alpha$ junction in the simulations. When the system reaches
equilibrium, the dimensionless chemical potential (energy unit is
chosen to be $k_BT$) of type-$\alpha$ junctions
($\tilde\mu_{\alpha}$), free ligands ($\tilde\mu_{L\alpha}$), and
free receptors ($\tilde\mu_{R\alpha}$) satisfy the condition \BEA
\tilde\mu_{\alpha} = \tilde\mu_{R\alpha} + \tilde\mu_{L\alpha}. \EEA
The chemical potentials of the free ligands and receptors are
related to their densities by $\psi_{L\alpha} =
e^{\tilde\mu_{L\alpha}}$ and $\psi_{R\alpha} =
e^{\tilde\mu_{R\alpha}}$, and we assume $\tilde\mu_{R\alpha} =
\tilde\mu_{L\alpha}$. Therefore, the relations between $E_{eff+}$
and total density of junctions and free ligands and receptors are
shown in Tab. \ref{tab:density}. It is clear that the density
increase as the effective binding energy increases. For higher total
effective binding energy ( $E_{eff+}$), the total junction density
is higher. It is because the effective binging energy is the sum of
the binding energy and the entropy lost due to the formation of
ligand-receptor pair (junction). For larger effective binging
energy, the ligand and receptor prefer to bind to each other. In
biological systems, typical values of the number of ligands
($N_{L\alpha}$) and receptors ($N_{L\alpha}$) are on the order
$10^5$ per cell, where the area of the cell ($A_c$) $\sim$ $10-10^4$
$\mu m^2$. In general case, we except that total number of free
type-$\alpha$ ligands and receptors are the same order as
$N_{L\alpha}$ and $N_{R\alpha}$. Thus typical values of
$\psi_{L\alpha}$ and $\psi_{R\alpha}$ should be
$\mathcal{O}(10^{-2})$, and typical binding energy $E_{B\alpha}$
$\sim$ 10-20 $k_BT$, typical $E_{eff1}$ and $E_{eff2}$ should be of
order unity, and they can be positive of negative. As shown in Tab.
\ref{tab:density}, our choice of parameters correspond to typical
biological systems.

\begin{figure}[tbp]
\epsfxsize= 6 in \epsfbox{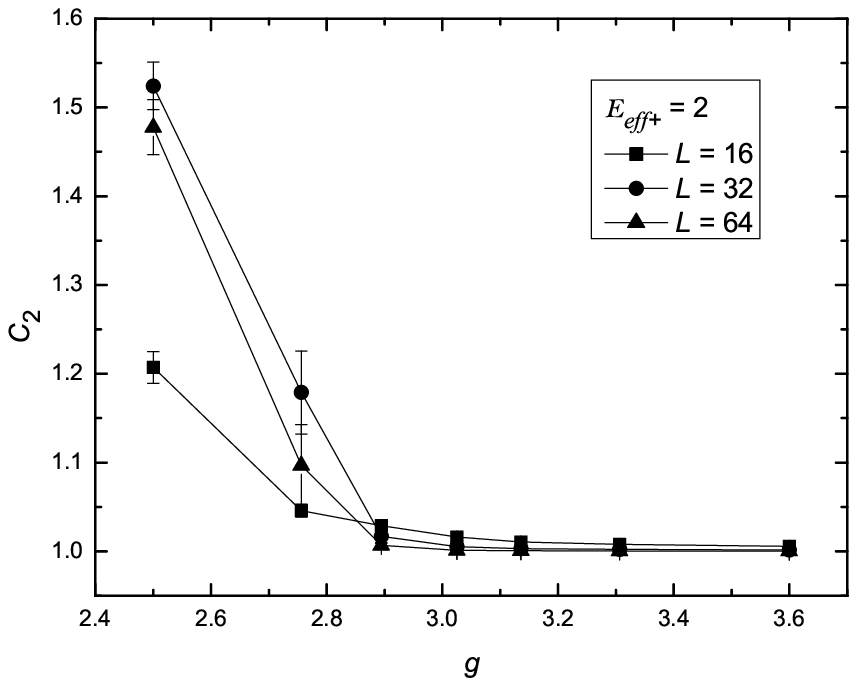} \vskip 24 pt \caption {$C_2$
as a function of $g$ for $L=16$ (square), 32 (circle),
and 64 (triangle) with parameters $\Lambda_1 =\Lambda_2 =
0.2$, $E_{eff+} = 2$ ($E_{eff1} = E_{eff2} = 1$), and $\ell_1 = 15$.
The common intersection point in this case is near $g=2.8$.} \vskip
30 pt \label{fig:Eeff1_C2}
\end{figure}

\begin{figure}[tbp]
\epsfxsize= 6 in \epsfbox{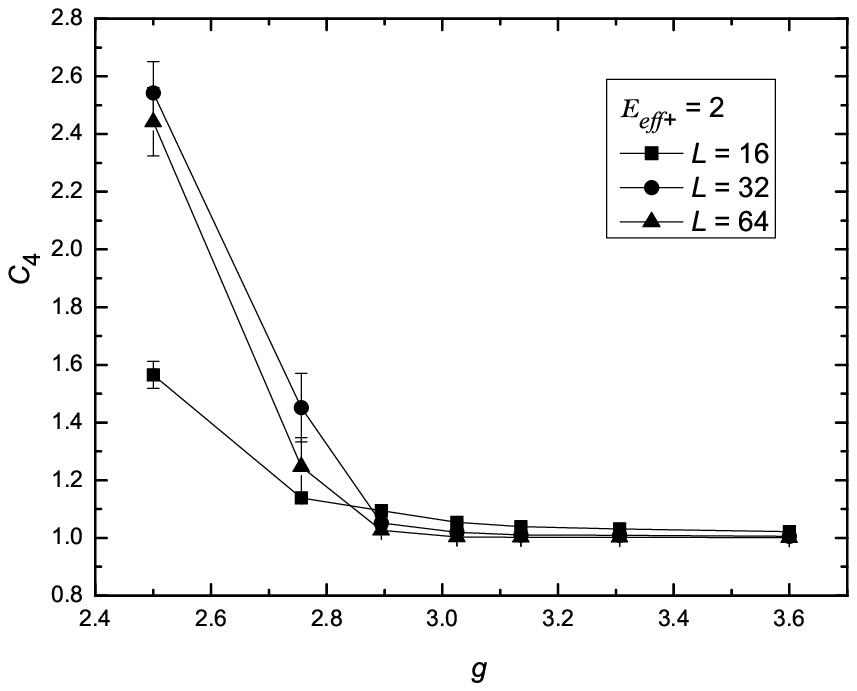} \vskip 24 pt \caption {$C_4$
as a function of $g$ for $L=16$ (square), 32 (circle),
and 64 (triangle) with parameters $\Lambda_1 =\Lambda_2 =
0.2$, $E_{eff+} = 2$ ($E_{eff1} = E_{eff2} = 1$), and $\ell_1 = 15$.
The common intersection point in this case is near $g=2.8$.} \vskip
30 pt \label{fig:Eeff1_C4}
\end{figure}

\begin{figure}[tbp]
\epsfxsize= 5.5 in \epsfbox{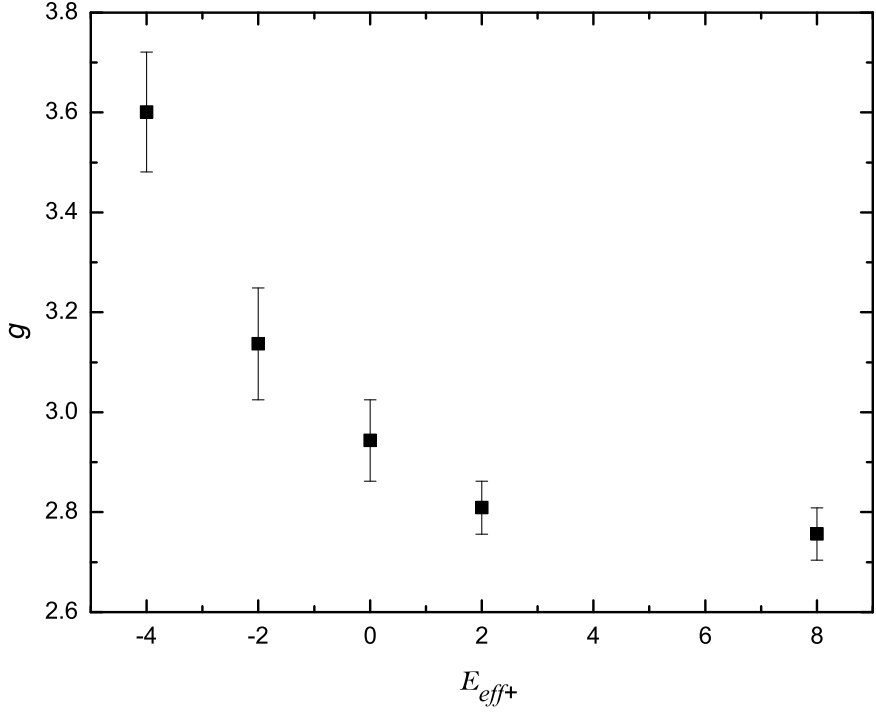} \vskip 24 pt \caption
{The critical value of $g$ as a function of $E_{eff+}$ for
$E_{eff-}=0$, $\Lambda_+ = 0.4$, $\lambda=0$, and $\ell_1=15$. The
critical value of $g$ decreases with the increase of $E_{eff+}$,
this is because the fluctuation of the membrane is smaller when the
junction density is higher. The data points in the figure are
partial result of our simulations. It is because the statistical
error increases as the value of $E_{eff+}$ decreases. Thus, we show
the data points for $-4\leq E_{eff+}\leq 8$.} \vskip 30 pt
\label{fig:critical g}
\end{figure}

Typical $C_2$ (Fig. \ref{fig:Eeff1_C2}) and $C_4$ (Fig.
\ref{fig:Eeff1_C4}) diagrams for several values of $L$ are shown.
From Fig. \ref{fig:Eeff1_C2} and \ref{fig:Eeff1_C4}, we find that
$g_c\simeq2.8$. We can estimate critical point in this way for
different $E_{eff+}$. Therefore, we plot the critical values of $g$
with respect to $E_{eff+}$ in Fig. \ref{fig:critical g}. It shows
that the critical value of $g$ decreases with the increase of
$E_{eff+}$, this is because the fluctuation of the membrane is smaller
when the junction density is higher.
\begin{table}[tbp]
\caption{Total density of junctions and free ligands and receptors}
\vskip 24 pt
\begin{center}
\begin{tabular}{ccc}
\hline \\
$E_{eff+}$ & junctions(\%) & free ligands and receptors(\%) \\ \\
\hline\hline
-8    &  $\thicksim$   1.6 & $\thicksim$  $2\times10^{-4}$ \\
-6    &  $\thicksim$   4.8 & $\thicksim$  $5\times10^{-3}$ \\
-4    &  $\thicksim$  11.2 & $\thicksim$  $10^{-3}$ \\
-2    &  $\thicksim$  28   & $\thicksim$  $3\times10^{-3}$ \\
 0    &  $\thicksim$  50   & $\thicksim$  $9\times10^{-3}$ \\
 2    &  $\thicksim$  73   & $\thicksim$  $0.03$ \\
 4    &  $\thicksim$  87   & $\thicksim$  $0.07$ \\
 6    &  $\thicksim$  95   & $\thicksim$  $0.2$ \\
 8    &  $\thicksim$  98   & $\thicksim$  $0.5$ \\
\hline
\end{tabular}%
\end{center}
\vskip 36 pt \label{tab:density}
\end{table}

\chapter{Conclusions}
\begin{figure}[tbp]
\epsfxsize= 5.5 in \epsfbox{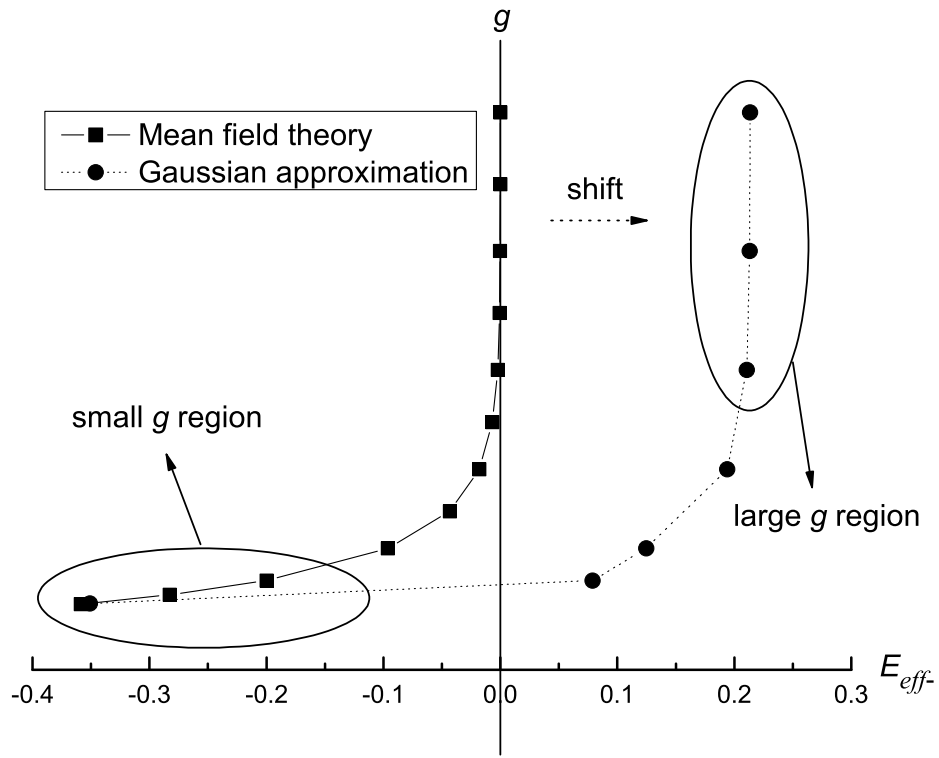} \vskip 24 pt \caption {The schematic
phase diagram for mean field theory (solid line) and Gaussian
approximation (dashed line).} \vskip 30 pt \label{fig:summary}
\end{figure}
In this thesis, we have studied the adhesion-induced phase
separation of multi-component membranes by theoretical and numerical
methods. In this system, two different types of junctions (they are
different in the natural length and the flexibility) mediate the
membrane adhesion. The phase diagram for a system with $\lambda>0$
is showen schematically in Fig. \ref{fig:summary}. Three methods are
used to study the phase
separation of this system. \\
1. In mean field theory, we integrate over all distributions of
junctions in the partition function and obtain an effective membrane
potential. The junction height difference and the junction
flexibility difference are two important factors that affect the
phase behavior. We discuss the phase diagrams in both symmetric case
(both types of junctions have the same flexibility but different
natural length) and asymmetric case (different type of junctions
have different flexibilities and natural length). After analyzing
the properties of the potential, we find that in symmetric case, the
phase separation is driven by the height difference of the
junctions. In the region of sufficiently large height difference in
the asymmetric case, the junction height difference is the main
factor that drives the phase separation. However, in the small
height difference region in the asymmetric case, the minimum energy
of the more rigid junctions is ``lowered'' by the binding energy of
the softer junctions. As a result, in this region the phase
coexistence occurs when the effective binding energy of the more
rigid (softer) junctions are smaller (larger). Thus, the phase
boundary moves toward smaller $E_{eff-}$.

2. Gaussian approximation is used to study the fluctuation effect on
the phase coexistence curves of the asymmetric case. When the
junction height difference is large, because the softer junctions
can be stretched or compressed easier than the more rigid junctions
and have higher entropy, comparing to the mean field theory, the
phase coexistence curves move to higher effective binding energy of
the more rigid junctions. Furthermore, the softer the junctions are,
the higher the the entropy they have. Thus, the phase boundary shift
is more significant for system with larger junction flexibility
difference and smaller total junction effective binding energy. In
small junction height difference region, the phase boundary moves
toward the mean-field critical point as the junction height
difference decreases (because Gaussian theory and mean field theory
predict the same critical point).

3. Monte Carlo simulations simulate the symmetric case to study the
effect of junction density on the critical point. We find that the
junction density increases when the total effective binding energy
increases. Meanwhile, the phase separation occurs at larger junction
height difference when junction density of the system is lower.
Because the fluctuation of membrane is smaller when the junction
density is higher.

Experimentally, the transition from shorter-junction-rich phase to
longer-junction-rich phase can be induced by applying a mechanical
pulling force on the system. This is because the effective binding
energy of the longer junctions increases and that of the shorter
junctions decrease under a pulling force. Thus, the phase transition
can be observed in typical experiments.

In summary, we have provided a general physical picture for the
adhesion-induced phase separation of multi-component membranes. In
our theory, we assume that the membranes bind to each other in the
whole process. It is possible to study binding/unbinding transition
in the future. In this case, the membrane-membrane collisions may
drive a different kind of phase separation. Another possible future
work is to study the dynamics of adhesion and detachment of this
type of systems.

\newpage
{\thispagestyle{empty} \topskip 3in
\begin{center}
{\Huge \textbf{APPENDICES}}
\end{center}}

\appendix
\chapter{Elasticity of two fluctuating membranes}
This appendix shows the calculation that reduces a two-membrane
Hamiltonian to a single-membrane Hamiltonian. If we consider only
the bending energy of the membranes then the Hamiltonian of the
system with two membranes in the solvent can be written as \BEA H =
\int d^{2}r \left\{ \frac{\kappa_1}{2} \left[
\nabla^{2}z_1(\mathbf{r}) \right]^{2} + \frac{\kappa_2}{2} \left[
\nabla^{2}z_2(\mathbf{r}) \right]^{2} \right\}, \label{eq:app. H
with two membranes} \EEA where $z_1$ and $z_2$ are the height of the
upper membrane and the lower membrane from the reference plane,
respectively. Define $u=z_1+z_2$ and $h=z_1-z_2$, thus
$z_1=\frac{1}{2}(u-h)$ and $z_2=\frac{1}{2}(u+h)$. Substitute
$[\nabla^{2}z_1]^2 = [\frac{1}{2}(\nabla^{2}u-\nabla^{2}h)]^2$ and
$[\nabla^{2}z_1]^2= [\frac{1}{2}(\nabla^{2}u-\nabla^{2}h)]^2$ into
Eq. (\ref{eq:app. H
with two membranes}) leads to %
\BEA H &=& \frac{1}{8} \int d^{2}r \left\{
(\kappa_1+\kappa_2) \left[ \nabla^{2}u(\mathbf{r}) \right]^{2} +
(\kappa_1+\kappa_2) \left[ \nabla^{2}h(\mathbf{r}) \right]^{2}
\right. \nonumber \\ && \left. +\ 2(\kappa_1-\kappa_2)
\nabla^{2}u(\mathbf{r})\nabla^{2}h(\mathbf{r}) \right\} \nonumber \\
&=& \int d^{2}r \left\{ \frac{1}{2}
\frac{\kappa_1\kappa_2}{\kappa_1+\kappa_2} \left[
\nabla^{2}h(\mathbf{r}) \right]^{2} + \frac{\kappa_1+\kappa_2}{8}
\left[ \nabla^{2}u(\mathbf{r}) \right. \right. \nonumber \\ &&
\left. \left. + \frac{\kappa_1-\kappa_2}{\kappa_1+\kappa_2}
\nabla^{2}h(\mathbf{r}) \right]^{2} \right\}
\label{eq:app. H with one membrane}\EEA %
We integrate out the second term in Eq. (\ref{eq:app. H with one
membrane}) and define $\kappa =
\frac{\kappa_1\kappa_2}{\kappa_1+\kappa_2}$, then the Hamiltonian
becomes \BEA H = \int d^{2}r \left\{ \frac{\kappa}{2} \left[
\nabla^{2}h(\mathbf{r}) \right]^{2} \right\}. \label{eq:app. H after
reduction}\EEA Similarly, for systems with non-vanishing surface
tension, \BEA H = \int d^{2}r \left\{ \frac{\kappa}{2} \left[
\nabla^{2}h(\mathbf{r}) \right]^2 + \frac{\gamma}{2} \left[ \nabla
h(\mathbf{r}) \right]^2 \right\}, \label{eq:app. H final type}\EEA
where $\gamma=\frac{\gamma_1\gamma_2}{\gamma_1+\gamma_2}$.

\chapter{Nondimensionalization of the Hamiltonian}
The Hamiltonian of the system is \BEA H &=& \int dxdy \left\{
\frac{\kappa}{2} \left[ \nabla^{2}h(x,y) \right]^{2} + \frac{\gamma
}{2} \left[ \nabla h(x,y) \right] ^{2} \right. \nonumber \\ &&
\left. + \sum_{\alpha=1}^{2}\Phi _{\alpha }(x,y)\frac{\lambda
_{\alpha }}{2} \left[ h(x,y) - h_{\alpha } \right] ^{2} -
\sum_{\alpha =1}^{2}\Phi_{\alpha }E_{B\alpha}\right\}.
\label{eq:app. H}
\EEA %
The unit length in $z$-direction is $h_{0}= a/\sqrt{\frac{\kappa
}{k_{B}T}}$. The discretized Laplacian and gradient of $h$ in two
dimensional space are \BEA \nabla^2 h &=& \frac{\partial^2
h}{\partial x^2}
+ \frac{\partial^2 h}{\partial y^2} \nonumber \\
&=&  \frac{\partial}{\partial x}( \frac{h_{x + \frac{a}{2}, y} -
h_{x - \frac{a}{2}, y }}{a} ) + \frac{\partial}{\partial
y}(\frac{h_{x, y + \frac{a}{2}} - h_{x,
y - \frac{a}{2}}}{a} ) \nonumber \\
&=& \frac{1}{a^2}( h_{x+a,y} + h_{x-a,y} + h_{x,y+a}+ h_{x,y-a}  - h_{x,y}) \nonumber \\
&=& \frac{1}{a} \sqrt{\frac{k_{B}T}{\kappa }} (\ell
_{x+a,y}+\ell_{x-a,y}+\ell_{x,y-a}+\ell _{x,y+a}-4\ell_{x,y})
\nonumber \\
&\equiv& \frac{1}{a} \sqrt{\frac{k_{B}T}{\kappa}} \Delta _{d}\ell
_{x,y} = \frac{1}{a} \sqrt{\frac{k_{B}T}{\kappa}}\Delta _{d}\ell
_{i} \label{eq:app. Laplacian}\EEA %
and %
\BEA \nabla h &=& \frac{\partial h}{\partial x} \hat{\textbf{i}}
+ \frac{\partial h}{\partial y} \hat{\textbf{j}} \nonumber \\
&=& \frac{1}{2a} \left[ ( h_{x+a,y} - h_{x-a,y} ) \hat{\textbf{i}} +
( h_{x,y+a} - h_{x,y-a} ) \hat{\textbf{j}} \right] \nonumber \\
&=& \frac{1}{2a} \sqrt{\frac{k_{B}T}{\kappa}} \left[ (\ell_{x+a,y} -
\ell_{x-a,y}) \hat{\textbf{x}} + (\ell_{x,y+a}-\ell_{x,y-a})
\hat{\textbf{y}} \right] \nonumber \\
&\equiv& \frac{1}{2a} \sqrt{\frac{k_{B}T}{\kappa}} \nabla \ell
_{x,y} = \frac{1}{2a} \sqrt{\frac{k_{B}T}{\kappa}} \nabla \ell_{i}.
\label{eq:app. gradient} \EEA %
Also defined here is $\phi_{\alpha }(x,y) = a^{2}\Phi _{\alpha
}(\mathbf{r})$, the number of type-$\alpha$ junctions at site
$(x,y)$. Hence, the non-dimensionalized Hamiltonian of the system is
\BEA \tilde H &=& \sum _{i = 1 }^{N} \left\{ \frac{1}{2}(\Delta _d
\ell _i)^2 + \frac{1}{2} \frac{\gamma a^2}{\kappa}(\mathbf{\nabla}
\ell _i)^2 \nonumber \right. \\ && \left. + \sum _{\alpha = 1}^{2}
\phi _{\alpha}(i) \left[ \frac{1}{2} \frac{\lambda _{\alpha} a
^2}{\kappa} \left[ \ell(i) - \ell _{\alpha} \right]^2 - \tilde E
_{B\alpha} \right] \right\}. \label{eq:app. non-dim H}
\EEA%

\newpage
{\thispagestyle{empty} \topskip 3in
\begin{center}
\Huge \bf BIBLIOGRAPHY
\end{center}}

\bigskip


\begin{thebibliography}{99}
\bibitem{ref:BPG_93} R. Bruinsma, M. Goulian, and P. Pincus,
   Biophys. J. {\bf 67}, 746, (1994).
\bibitem{ref:bruinsma_PRL_95} D. Zuckerman and R. Bruinsma,
   Phys. Rev. Lett., {\bf 74}, 3900, (1995).
\bibitem{ref:lipowsky_book_95} R. Lipowsky and E. Sackmann,
   {\it The Structure and Dynamics of Membranes}, (Elsevier, Amsterdam, 1995).
\bibitem{ref:lipowsky_PRL_96} R. Lipowsky,
   Phys. Rev. Lett., {\bf 77}, 1652, (1996).
\bibitem{ref:sackmann_BJ_97} A. Albersd\"{o}rfer, T. Feder, and E. Sackmann,
   Biophys. J., {\bf 73}, 245, (1997).
\bibitem{ref:sackmann_EPL_97} J. Nardi, T. Feder, and E. Sackmann,
   Europhys. Lett. {\bf 37}, 371, (1997).
\bibitem{ref:sackmann_PRE_00} R. Bruinsma, A. Behrisch, and E. Sackmann,
   Phys. Rev. E, {\bf 61}, 4253, (2000).
\bibitem{ref:lipowsky_PRE_00} T. R. Weikl, R. R. Netz, and R. Lipowsky,
   Phys. Rev. E, {\bf 62} R45, (2000).
\bibitem{ref:andelman_EPE_00} S. Komura and D. Andelman,
   Eur. Phys. J. E, {\bf 3} 259, (2000).
\bibitem{ref:lipowsky_PRE_01} T. R. Weikl and R. Lipowsky,
   Phys. Rev. E, {\bf 64}, 011903, (2001).
\bibitem{ref:Burroughs_BPJ_02} N.J. Burroughs and C. W\"{u}lfing,
   Biophys. J. {\bf 83}, 1784, (2002).

\bibitem{ref:exp} C. R. F. Monks, et. al., Nature, {\bf 395}, 82, (1998),
                  G. Grakoui, et. al., Science, {\bf 285}, 221, (1999),
                  and D. M. Davis, et. al.,
   Proc. Natl. Acad. Sci. USA, {\bf 96}, 15062, (1999).
\bibitem{ref:alberts_book} B. Alberts, D. Bray, J. Lewis, M. Faff, K. Roberts, and J. D. Watson,
   {\it Molecular Biology of the Cell}, 3rd ed. (Garland, New York, 1994).
\bibitem{ref:Qi_pnas_01} See, S.Y. Qi, J. T. Groves, and A. K. Chakraborty,
   Proc. Natl. Acad. Sci. USA, {\bf 98}, 6548, (2001).
\bibitem{ref:Raychaudhuri_PRL_03} S. Raychaudhuri, A. K. Chakraborty, and M. Kardar,
   Phys. Rev. Lett. {\bf 91}, 208101, (2003).
\bibitem{ref:lipowsky_EPL_02} T.R. Weikl, J.T. Groves, and R. Lipowsky,
   Europhys. Lett. {\bf 59}, 916 (2002).
\bibitem{ref:lipowsky_BPJ_04} T.R. Weikl and R. Lipowsky,
   Biophys. J. {\bf 87}, 3665, (2004).
\bibitem{ref:Chen_PRE_03} H. Y. Chen, Phys. Rev. E, {\bf 67}, 031919,(2003).
\bibitem{ref:sackmann_BJ_95} H. Strey, M. Peterson, and E. Sackmann,
   Biophys. J. {\bf 69}, 478, (1995).
\bibitem{ref:hochmuth_BPJ_92} Needham. D. and R. M. Hochmuth,
   Biophys. J. {\bf 61}, 1664, (1992).
\bibitem{ref:bell_BPJ_84} G.I. Bell, M. Dembo, and P. Bongrand,
   Biophys. J. {\bf 45}, 1051, (1984).
\bibitem{ref:lipowsky_PRL_99} R. Geotz, G. Gompper, and R. Lipowsky,
   Phys. Rev. Lett., {\bf 82}, 221, (1999).
\bibitem{ref:binder_book} K. Binder and D. W. Heerman,
        {\it Monte Carlo Simulation in Statistical Physics An Introduction},
        2nd corrected ed. (Berlin, New York, Springer-Verlag, 1992).
\bibitem{ref:andelman_EPE_02}T.R. Weikl, D. Andelman, S. Komura, and R. Lipowsky,
   Eur. Phys. J. E, {\bf 8}, 59, (2002).
\end{thebibliography}
\end{document}